\documentclass[journal=jacsat,manuscript=article]{achemso}

\usepackage[version=3]{mhchem} 
\usepackage{newunicodechar}
\usepackage{amsmath}
\newunicodechar{∼}{\textasciitilde}

\author{Rishabh Dey}
\email{rishabh1dey@gmail.com}
\author{Salvina Sharipova}
\author{Konstantin Popov}
\affiliation[University of North Carolina]
{University of North Carolina at Chapel Hill - Eshelman School of Pharmacy}

\title[An \textsf{achemso} demo]
  {All-atomistic Transferable Neural Potentials for Protein Solvation}

\abbreviations{IR,NMR,UV}

\begin{document}

\begin{tocentry}

Some journals require a graphical entry for the Table of Contents.
This should be laid out ``print ready'' so that the sizing of the
text is correct.

Inside the \texttt{tocentry} environment, the font used is Helvetica
8\,pt, as required by \emph{Journal of the American Chemical
Society}.

The surrounding frame is 9\,cm by 3.5\,cm, which is the maximum
permitted for  \emph{Journal of the American Chemical Society}
graphical table of content entries. The box will not resize if the
content is too big: instead it will overflow the edge of the box.

This box and the associated title will always be printed on a
separate page at the end of the document.

\end{tocentry}

\begin{abstract}
  Implicit solvent models are widely used to decrease the number of solvent degrees of freedom and enable the calculation of solvation energetics without water molecules. However, its accuracy often falls short compared to explicit models. Recent advancements in neural potentials have shown promise in drug discovery, but transferability remains a persistent challenge. Here, we introduce the Protein Hydration Neural Network (PHNN), an implicit solvent model that extends analytical continuum solvation by learning transferable corrections to model parameters instead of applying post hoc adjustments to final energies. The model is explicitly designed to maximize data efficiency by leveraging physical priors embedded in the data. We demonstrate that PHNN improves accuracy relative to traditional analytical methods and maintains predictive accuracy on out-of-domain protein systems.
\end{abstract}
\newpage
\section{Introduction}

Accurate conformational sampling of biomolecules is required to simulate various chemical processes and is imperative for structural analysis and drug discovery \cite{Zuckerman2011}. Molecular dynamics (MD) simulations typically require explicit water molecules to calculate solvent-solute interactions and provide accurate predictions. In contrast, implicit solvent models, or solvation continuum models, treat the solvent as a dielectric continuum, greatly reducing degrees of freedom of solvation and computational cost \cite{Huang2007}. Popular examples are Poisson–Boltzmann (PB) or Generalized Born (GB) models \cite{Onufriev2019, Kleinjung2014}. 

Traditional implicit solvent models suffer from several fundamental limitations. The accuracy of these models are limited as they fail to capture instantaneous solvent fluctuations and specific solvent-solute interactions. The total solvation energy of a system can be represented as the sum of the polar and nonpolar components:

\begin{equation}
    \Delta G_{\text{solv}} = \Delta G_{\text{polar}} + \Delta G_{\text{nonpolar}}
\end{equation}

While GB and PB may provide an accurate assessment of the polar component, a major limitation is the oversimplification of the nonpolar component to a simple solvent-accessible surface area (SASA) term. A popular methodology utilizing this is Generalized Born Neck 2 (GBn2), where the model utilizes the effective Born radii, \textit{$R_i$}, a representation of how buried the atom is in the molecule \cite{nguyen2013}:

\begin{equation}
    R_i = (\rho_i^{-1} - I_i)^{-1}
\end{equation}
Where $\rho$ represents the intrinsic radius, and \textit{I} represents the Coulomb integral:
\begin{equation}
    I_i = \frac{1}{4\pi} \int_{\Omega, r > \rho_i} \frac{1}{r^4} d^3r.
\end{equation}
GBn2 then calculates the SASA with the following equations:
\begin{equation}
    A = \sum_i^N 4\pi (r_i + r_p)^2 \cdot \left( \frac{r_i}{R_i} \right)^6
\end{equation}
\begin{equation}
\Delta G_{\text{nonpolar}} = \sum_i^N \sigma_i \cdot A_i
\end{equation}

While there are several methods for the calculation of the SASA component, the fundamental idea has many theoretical drawbacks (e.g, overstabilization of non-polar pairwise interactions)\cite{wagoner2006}. Nevertheless, erroneous non-polar interactions are allowed to exist as they are typically smaller than polar contributions.

However, the polar contributions also have accuracy issues. Previous studies have observed supralinear electrostatic solvent responses with volatile solvent-solute electrostatic interactions\cite{muddana2013}. Muddana et al. states, “accounting for solute electrostriction may be a productive approach to improving the accuracy of continuum solvation models." We can model the electrostatic solvent responses with an ABPS map as seen in Figure 1; highly charged areas can warp the surrounding dielectric \cite{baker2001apbs}. The GB analytical formula:
\begin{equation}
\Delta G_{\text{GB}} = -\frac{1}{2} \left( \frac{1}{\epsilon_{in}} - \frac{1}{\epsilon_{out}} \right) \sum_{i,j} \frac{q_i q_j}{f_{GB}}
\end{equation}
\begin{equation}
f_{GB} = \sqrt{r_{ij}^2 + R_i R_j \exp\left( \frac{-r_{ij}^2}{4 R_i R_j} \right)}
\end{equation}

The GB equation assumes a constant dielectric for all pairwise interactions in the MD simulation \cite{bashford2000}. However, the solute and solvent dielectric should change based on the electrostatic environment. Furthermore, the pairwise GB screening function, \textit{f}, underestimates dielectric shielding between two nearby charges because each atom's Born radius is computed independently \cite{nguyen2013}; the mutual desolvation of the ion pair is only partially captured. While GBn2 attempts to combat this, the issue persists. This interaction leads to a known issue where Generalized Born implicit solvent models poorly represent Glu/Lys interactions \cite{schiemann2022glulys}. 

\begin{figure}
    \centering
    \includegraphics[width=0.5\linewidth]{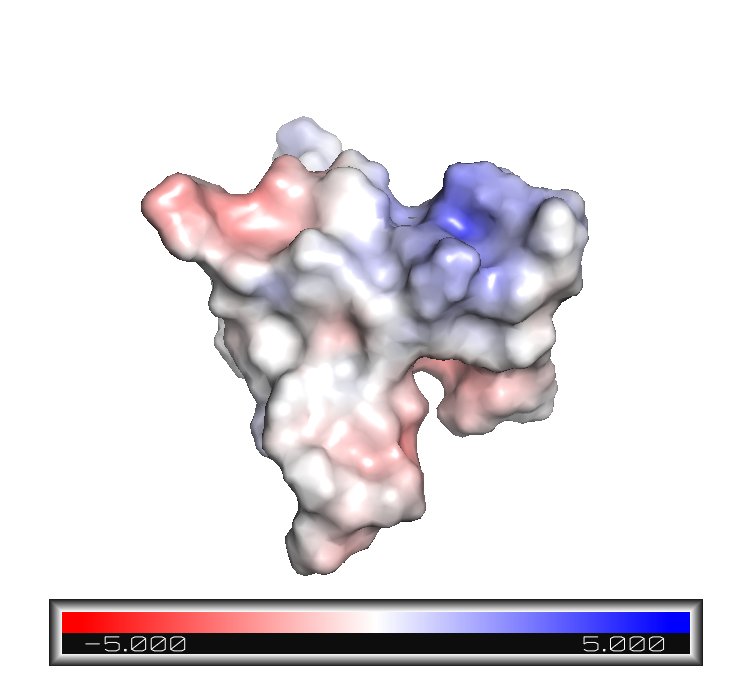}
    \caption{Adaptive Poisson-Boltzmann Solver (ABPS) electrostatic map of CATH domain 4ifdG01 (1821 atoms). As the electrostatic forces change between solvent-solute interactions, the dielectric should not be treated as a constant value. }
    \label{fig:placeholder}
\end{figure}

The GB approach also breaks at the protein surface where water fails to behave as a continuum. The first and second hydration shells have structured density peaks \cite{grudinin2023hydration}. Capturing the angular anisotropy of the solvation shell remains one of the most valuable potential additions to implicit solvent models. For instance, Tolokh et al.’s model, IWM-GB, explicitly accounted for the effects of multipole moments of water molecules in the first hydration shell of a solute, leading to improved accuracy \cite{tolokh2024iwmgb}. Additionally, the model accounts for polar-nonpolar coupling rather than the simple sum of the components, as a polar residue next to a hydrophobic patch has different solvation than the same residue in a fully aqueous environment.

 Utilizing machine learning, we can resolve many existing shortcomings by integrating the computational speed of implicit frameworks with the high-fidelity resolution of all-atom simulations \cite{molecules28104047}. However, while contemporary architectures have achieved transformative success in predicting static protein folds and sampling conformational ensembles, these models struggle to maintain transferability beyond their training distributions \cite{Chakravarty2024}. The result is a class of models that excel at capturing predictable protein conformational ensembles yet frequently neglect the energetic subtleties of solvent fluctuation. In drug discovery, where binding affinity and folding kinetics are paramount, physics-grounded potentials are essential. These models must prioritize fundamental interatomic forces to ensure predicted dynamics remain physically consistent rather than merely geometrically plausible.

Forcefield models solely predict energetics and utilize MD to maintain thermodynamic stability.  Models like ISSNet or AIMNet have been applied to learn solvation free energies, but transferability issues \cite{chen2021mlimplicit, zubatyuk2019aimnet}. When these models encounter sequences in the bioinformatic “twilight zone”, defined by $<30\%$ sequence identity, accuracy degrades sharply\cite{rost1999twilight}. At this threshold, models like CGSchNet overstabilize collapsed states; they lack the physical anchors required to generalize to disordered regions or novel folds underrepresented in the training set \cite{husic2020coarsegraining}.

Traditionally, training a forcefield model requires a force-matching paradigm.\cite{wang2019dynamics} This is mathematically defined by the loss function:
\begin{equation}
    \mathcal{L} = \left( \frac{\partial U}{\partial r_i} - \frac{\partial f}{\partial r_i} \right)^2
\end{equation}
where U represents the true potential energy of the solvated system and f represents the learned potential. The direct identity of implicit solvent models is to differentiate a hydration energy approximation to calculate the mean solvation force across all solvent conformations as represented below:
\begin{equation}
\Delta G_{\text{solv}} \approx   \left\langle \Delta G_{\text{implicit\_solv}}(\mathbf{r}) \right\rangle
\end{equation}
\begin{equation}
\frac{\partial \Delta G_{\text{solv}}}{\partial \mathbf{r}} = \left\langle \frac{\partial U}{\partial \mathbf{r}} \right\rangle
\end{equation}

We can train a model to remain consistent with the hydration free energy, or the potential of mean force (PMF), by either force-matching calculated mean forces or allowing the model to converge on instantaneous explicit forces. While force-matching is a powerful tool, pure neural potentials remain vulnerable to learning spurious correlations and unphysical shortcuts. As such, PHNN uses GBn2 as a backbone to force-match against solvation forces \cite{nguyen2013}. 

While post-hoc adjustment of GBn2 and existing implicit solvent models is not a novel approach, prior methods have constrained themselves to augmenting the output of the analytical framework, either by adding an energy correction term or by scaling Born radii, leaving the underlying physical variables of the equations unchanged \cite{katzberger2024gnn, dey2025lsnn, kramer2023gnn}. PHNN instead operates on the level of the equations themselves to correct the environment-dependent patterns that are parameterized. This richer correction mechanism prevents the model from redundantly learning existing physical information and establishes a methodology that extends beyond a delta-learning paradigm.

\section{Methods}

\subsection{Dataset}
The mdCATH dataset was used at a simulation temperature of 320 K, and only the initial simulation out of the 5 trials were considered, yielding a training set of approximately 2.1 million conformations \cite{mirarchi2024mdcath, orengo1997cath}. As the original dataset contains non-homologous domains at the S20 (20\%) sequence identity level, a randomized validation set of approximately 200,000 frames from separate protein domains was constructed. Dataset splits were preprocessed with OpenMM for the calculation of PHNN features (e.g., GBn2 parameters) \cite{nguyen2013, eastman2024openmm8}. Finally, 40 proteins were selected as an independent test set for the reported results. All reference simulations were done on OpenMM with CHARMM36 forcefield and TIP3P as explicit solvent \cite{jorgensen1983tip3p, best2012charmm36}. To sample beyond CATH domains, we generated a Ramachandran plot for alanine dipeptide \cite{ramachandran1963}.

\subsection{Model}

PHNN’s model, $f_{\theta}$, was built as a GBn2 correction model, utilizing intrinsic parameters as features for the network. While this procedure has been done before, key changes were made as displayed in Figure 2.
\begin{figure}[!h]
    \centering
    \includegraphics[width=1\linewidth]{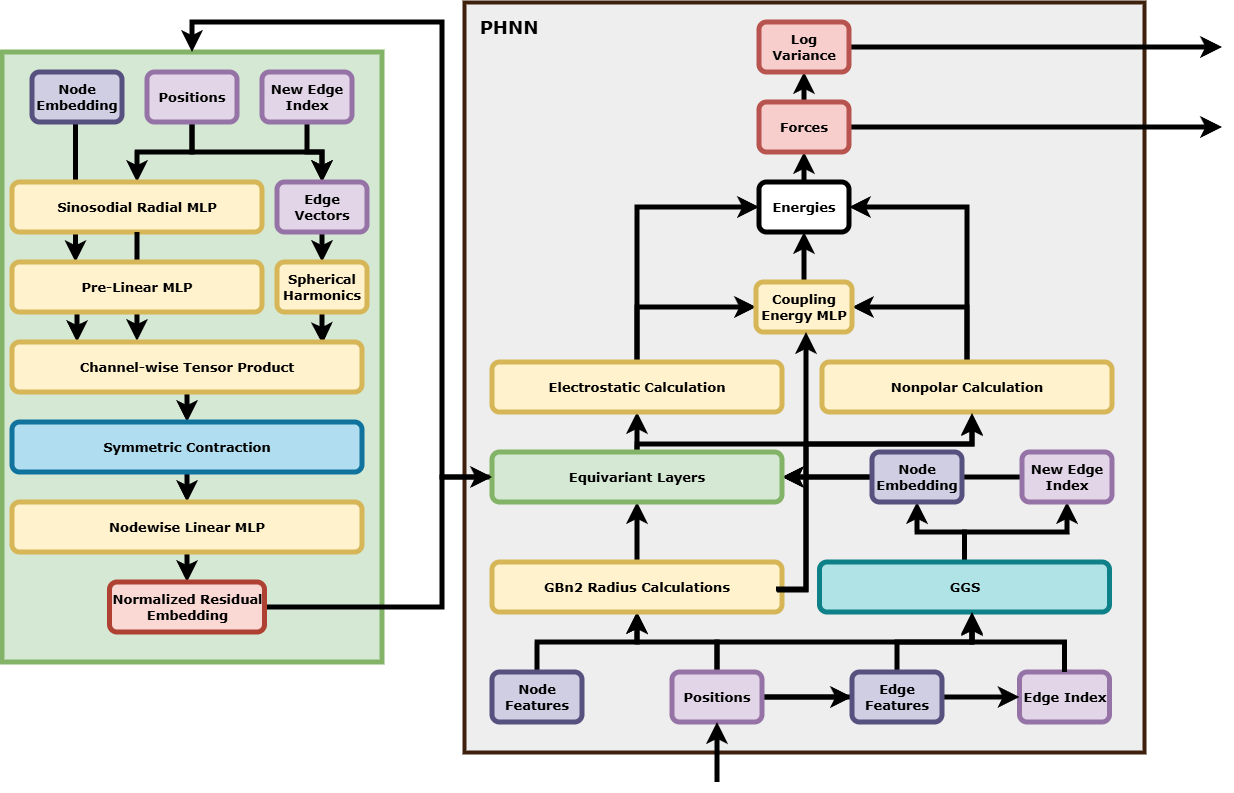}
    \caption{Overview of PHNN Model; PHNN takes in molecular dynamics information and outputs the respective forces that for that frame. Intermediate embeddings are calculated via E(3) layers and modulate the GBn2 calculation layers. }
    \label{fig:placeholder}
\end{figure}

The equivariant architecture within the model serve many specific purposes. The layers were modeled through a custom pseudo-MACE architecture built on CueEquivariance \cite{batatia2022mace, nvidia2024cuequivariance}; the primary change being that the Bessel radial function was replaced with a sinusoidal basis function with a Behler-style polynomial cutoff to stay consistent with previously tested invariant layers and reduce additional overhead \cite{behler2007parrinello}. The equivariant layers enable the network to represent multipole contributions with quadrupoles, which have previously been shown to enhance model accuracy \cite{batzner2022nequip, thurlemann2022multipoles, muddana2013}. Furthermore, they serve as the basis for the correction mechanism in the GB formula.

\begin{figure}
    \centering
    \includegraphics[width=0.75\linewidth]{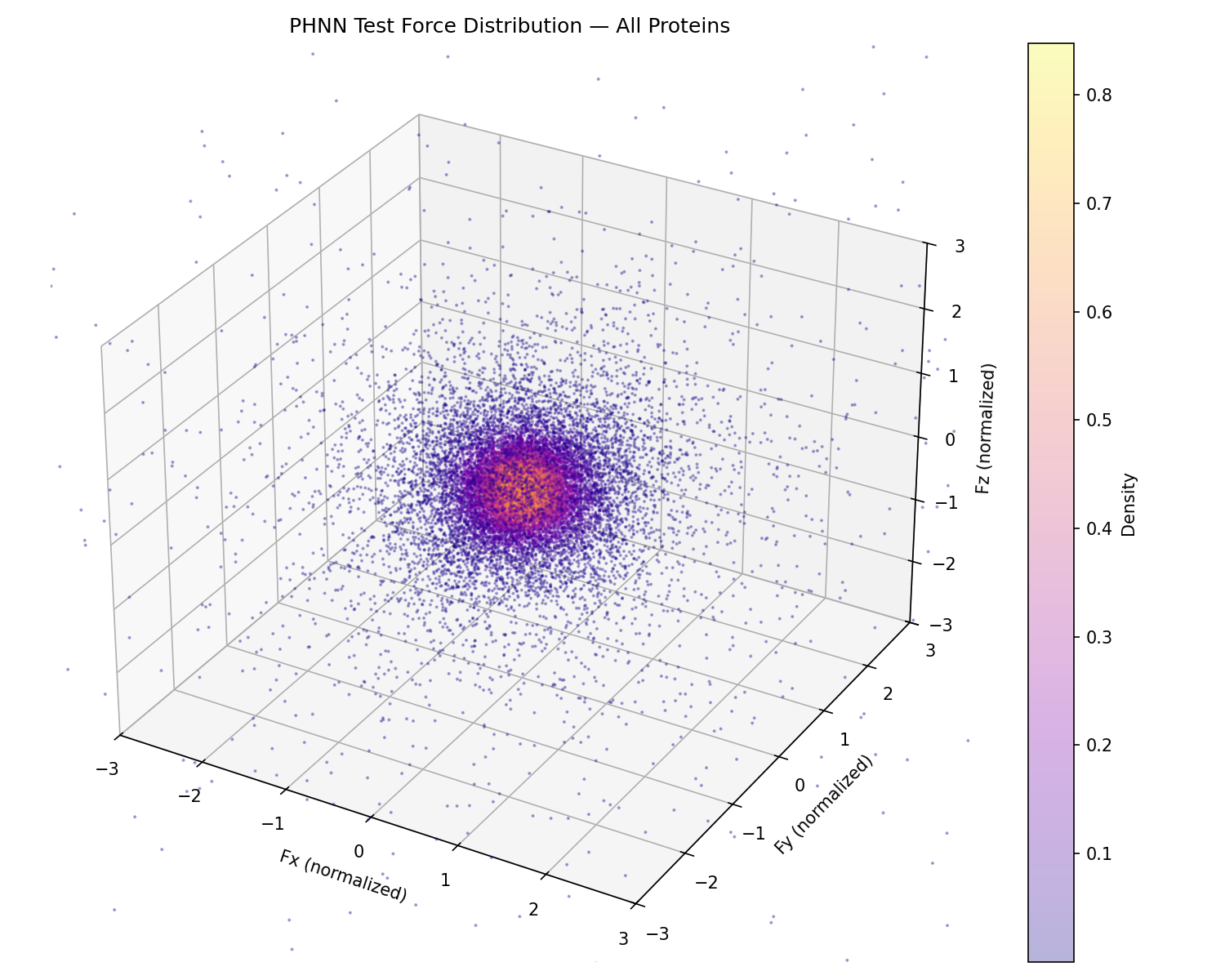}
    \caption{Standardized PHNN test set force distribution (n = 39 proteins). Forces natively forms a gaussian distribution.}
    \label{fig:placeholder}
\end{figure}

While PHNN attempts to generalize to the solvation mean forces, the mdCATH dataset contains instantaneous forces. However, with enough conformations modeled, the continuum can be captured in different regimes, and the instantaneous forcefield follows a Gaussian, as shown in Figure 3. To prevent overfitting, we employ a heteroscedastic loss function, defined as \cite{nix1994heteroscedastic}:

\begin{equation}
    \mathcal{L} = \frac{1}{2}\left(\frac{(y - \hat{y})^2}{\sigma^2} + \ln \sigma^2\right)
\end{equation}
To increase the stability of our model at high variances, we replaced the calculation of $\sigma^2$ with $\ln(\sigma^2)$, represented by $\nu$, and followed the $\beta$-NLL loss paradigm \cite{seitzer2022betanll}. The final PHNN loss function is: 
\begin{equation}
    \mathcal{L} = e^{\beta v} \cdot \frac{1}{2}\left(e^{-v}\sum_{d=1}^{3}(f_d - \hat{f}_d)^2 + v\right)
\end{equation}

A separate invariant GNN is used for variance estimation, taking as input the detached predicted forces, an equivariant variance latent parameter, and dominant GBn2 parameters.

\subsection{Tackling GBn2 limitations}

Higher-order equivariant representations allow us to capture the curvature and packing asymmetry of the atomic environment, properties that are directly relevant to nonpolar solvation and steric interactions. The nonpolar solvation energy is computed as a learned modulation of the SASA term. The surface tension coefficient $\gamma$ is also treated as a learnable parameter. The corrected nonpolar contribution is as follows:
\begin{equation}
    A_{i} = \gamma \cdot f_{\theta, nonpolar}\cdot (r_i + 0.14)^2
\end{equation}

This methodology has been applied before; as steric interactions decay quickly, they can easily be accounted for by the equivariant layers; the majority of direct corrections are related to electrostatic irregularities in the solvation continuum \cite{katzberger2024gnn}. Three atom-specific corrections were derived from the equivariant layers to improve GB electrostatics. Utilizing the equivariant layers, an atom-specific local solute dielectric was calculated to represent the polarizability of the protein interior around each atom. Moreover, a local solvent dielectric corrects the external screening environment. Salt-bridge stabilization can be addressed by modifying the screening function \textit{f}, which interpolates between the Born self-energy and classical Coulomb limits based on atomic separation relative to Born radii. An additional feed-forward network (FFN) modulates \textit{f} directly, taking as inputs the interacting electrostatic variables and learning environment-dependent corrections. Finally, a per-atom charge correction compensates for residual electrostriction effects between atoms. Together, these yield the corrected pairwise energy:
\begin{equation}
    q_i^* = q_i + \tanh(f_{\theta,q,i})
\end{equation}
\begin{equation}
    \epsilon_{k,i}^* = \alpha_k + \beta_k \cdot \sigma(f_{\theta, k, i}) \quad \text{for } k \in \{\text{solv, sol}\}
\end{equation}
where $\alpha_k$ and $\beta_k$ are bounds set to $\{20.0, 58.5\}$ for the solvent and $\{1.0, 19.0\}$ for the solute, respectively.
\begin{equation}
    \bar{\epsilon}_{k,ij} = \frac{1}{2}(\epsilon_{k,i}^* + \epsilon_{k,j}^*) \quad \text{for } k \in \{\text{solv, sol}\}
\end{equation}
\begin{equation}
    \Delta G_{\text{GB}} = \sum_{i<j} \left( \frac{1}{\bar{\epsilon}_{\text{sol},ij}} - \frac{1}{\bar{\epsilon}_{\text{solv},ij}} \right) \frac{q_i^* q_j^*}{f_{ij}^*}
\end{equation}
\begin{equation}
    f_{ij}^* = f_{ij} \left[ 1 + \lambda \cdot f_{\theta}(q_i^*, q_j^*, r_{ij}, B_i, B_j, \epsilon_{\text{solv},i}^*, \epsilon_{\text{sol},i}^*) \right]
\end{equation}

A polar-nonpolar coupling correction is applied via an MLP, taking as input previously determined values and calculated electrostatic and steric energies. Aggregated values are scaled by $\lambda_c$, initialized near zero for training stability. The total solvation energy per atom is then:
\begin{equation}
    \Delta G_i = \Delta G_{\text{polar},i} + \Delta G_{\text{nonpolar},i} + \Delta G_{\text{coupling},i}
\end{equation}

\subsection{Training protocol}
The model was trained for 2 Epochs with a batch size of 28, a learning rate (LR) of 5e-4, gradient clipping of 5.0, and a precision of bfloat16. Training time was 24 hours per epoch and approximately 49 hours in total on 4 L40 GPUs.

\section{Results}
\subsection{General force prediction}

\begin{figure} [!h]
    \centering
    \includegraphics[width=0.75\linewidth]{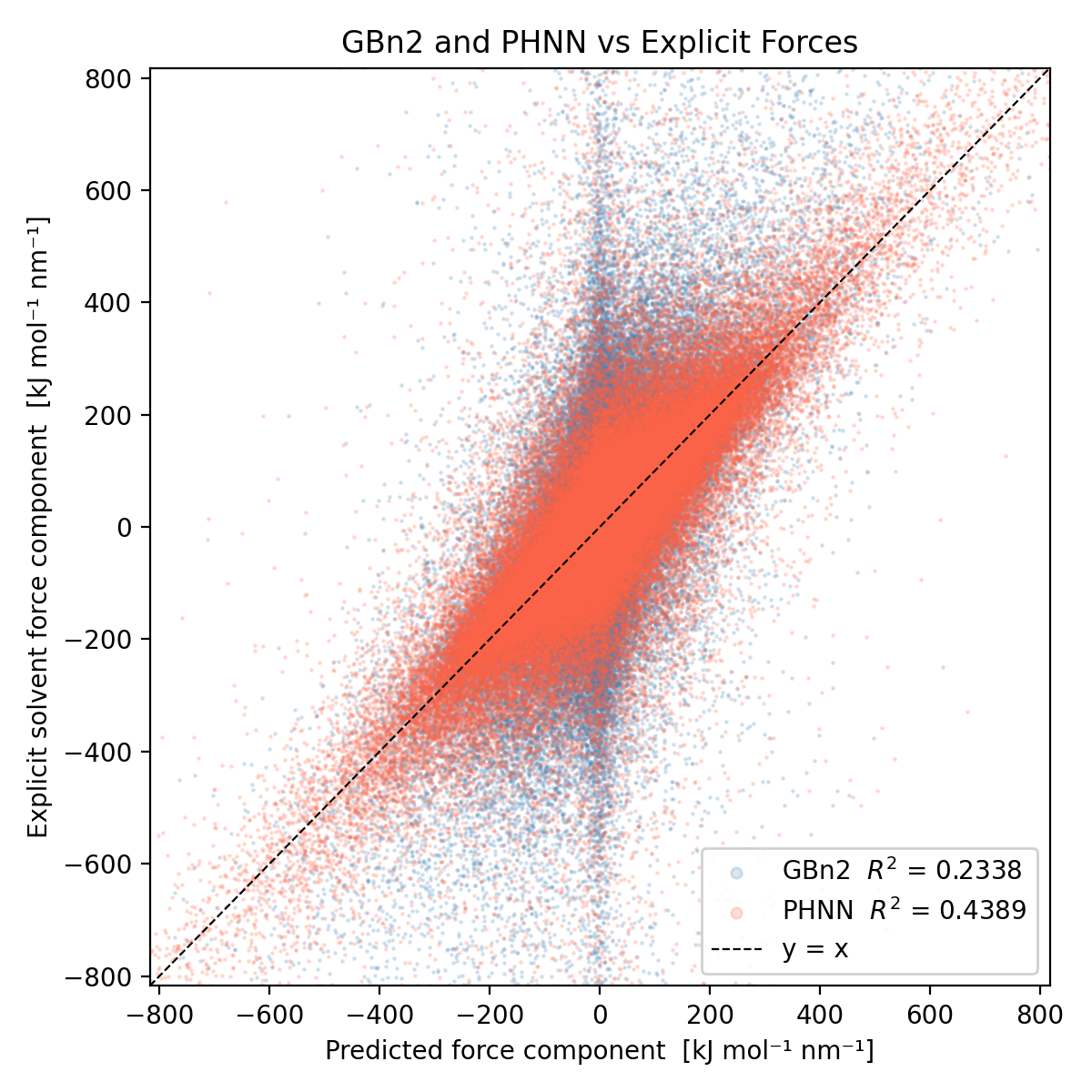}
    \caption{Atomistic force errors of Gbn2 (in blue) and PHNN (in orange) compared against TIP3P (n =39 proteins). }
    \label{fig:placeholder}
\end{figure}

Utilizing the OOD testing set, we calculated mean solvation forces on frozen equilibrated and nonequilibrated frames over 20 ns simulations on 39 proteins. PHNN achieves a mean MAE of  66.6 ± 9.4 kJ/(mol·nm) against explicit solvent forces, representing a 31.7\% reduction in error relative to GBn2 (97.5 ± 9.0 kJ/(mol·nm)). This improvement is consistent across proteins ranging from approximately 800 to 6000 atoms, with the standard deviation in PHNN MAE remaining comparable to that of GBn2. PHNN's MAE relative to GBn2 forces is larger than its MAE relative to explicit solvent forces, suggesting that PHNN may be making independent errors. 

An additional observation can be made from Figure 4: while PHNN and GBn2 forces reside in a similar space, GBn2 tends to underpredict solvation forces, setting them to 0. We suspect that the oversimplification of the nonpolar contribution is the primary contributor. 

However, there remains a clear flaw in our methodology; we note that the mean variance of explicit solvent forces across the test set is 6649 ± 946 (kJ/(mol·nm))², reflecting the inherently stochastic nature of instantaneous solvation forces and establishing a practical upper bound on the accuracy achievable by any deterministic implicit solvent model. This variance also implies that the reference explicit mean forces are themselves noisy estimates. In a secondary trial, PHNN achieves a mean MAE of 69.116 ± 13.041 kJ/(mol·nm) in comparison to GBn2’s 99.588 ± 12.314 kJ/(mol·nm)-- a 30.6\% improvement. As such, both GBn2 and PHNN may perform better than the raw numbers suggest.

\subsection{Free energy and simulation stability}

While the mean force convergence at single frames is meaningful, it does not represent dynamical stability or thermodynamic properties. As we lack precise values of solvation free energy, we instead represent the free energy of the system by calculating the density function of intrinsic free energies by running extended simulations. For this purpose, we ran 80 ns simulations in TIP3P, 25ns in GBn2, and 10ns in PHNN and calculated the root mean square deviation (RMSD) from native state, radius of gyration (ROG), and root mean square fluctuation (RMSF) every 100 frames on 4 proteins domains (798-5404 atoms). Simulation lengths were chosen based on the plateau of mean RMSD as a convergence criterion. After data collection, we utilized the Kernel Density Estimation function (KDE) to estimate the free energy landscape as displayed in Figure 5 \cite{rosenblatt1956kde}. 

\begin{figure}[!h]
    \centering
    \includegraphics[width=1\linewidth]{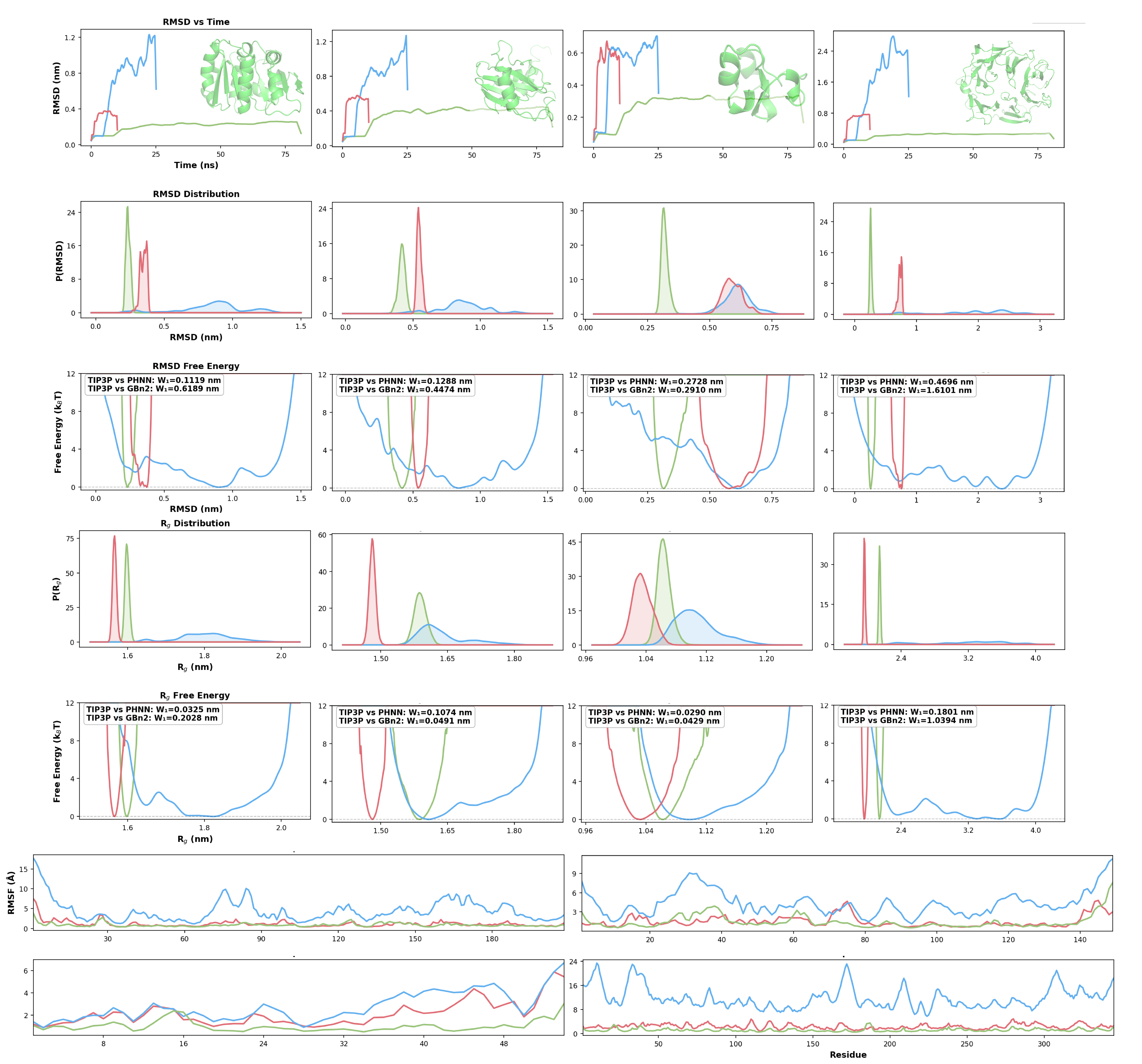}
    \caption{Dynamical stability and free energy analysis of four protein domains simulated under TIP3P (blue), PHNN (red), and GBn2 (green) solvent models. (Row 1) RMSD vs. time with ribbon representations of each protein domain (3al9A02, n=798 atoms; 2fz0A00, n=2387; 1imjA00, n=3128; 4bp9A02, n=5404). (Row 2) RMSD probability distribution relative to native structure. (Row 3) Free energy landscapes estimated via KDE of RMSD, with annotated W1 distance between TIP3P distribution and PHNN and TIP3P distribution and GBn2. (Row 4) ROG probability distribution. (Row 5) Free energy landscapes of ROG with annotated W1 values. (Row 6-7) Per-residue RMSF profiles across all four domains. Simulations were run on TIP3P (80 ns), GBn2 (25 ns), and PHNN (10 ns). The first 10 ns, 5 ns, and 1 ns of each simulation respectively was restrained for simulation stability.}
    \label{fig:placeholder}
\end{figure}

PHNN performed better overall across all domains, consistent with force prediction accuracy. Most notably is 4bp9A02 (n=5404 atoms) with a W1=0.46 and 1.61 nm for RMSD and W1= 0.18 and 1.03 with PHNN and GBn2 respectively. A common trend is that GBn2 continues to unfold, especially with larger protein domains, with RMSF peaking at 2.4 nm in 4bp9A02 and remaining elevated across all domains, suggesting persistent structural instability rather than localized flexibility. PHNN struggled more on smaller domains, 3al9A02 (n=798 atoms) and 2fz0A00 (n=2387 atoms), where RMSD drift for PHNN and GBn2 was nearly equivalent for the former and W1=0.10 and 0.04 nm for ROG in the latter. 

The performance gap between PHNN and GBn2 narrows at smaller scales because GBn2 parameters were calculated with explicit solvent dipeptides and small molecules. PHNN's broader dataset better accounts for irregular conformations not captured by GBn2's parameterization, and collectively, these results suggest PHNN offers more reliable implicit solvent dynamics for larger, structurally complex protein domains.

\begin{figure}[!h]
    \centering
    \includegraphics[width=1\linewidth]{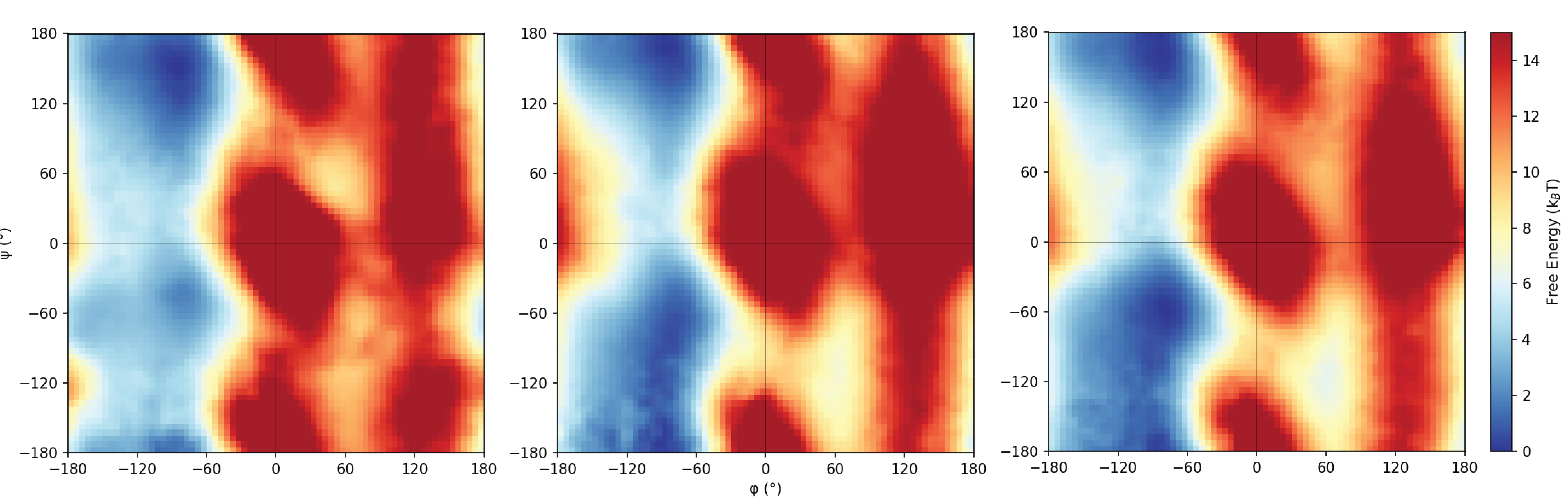}
    \caption{Ramachandran plots of alaline dipeptide of GBn2 (left), PHNN (middle), and TIP3P (right)}
    \label{fig:placeholder}
\end{figure}

To calculate PHNN's limits, we utilized alanine dipeptide. As the average residue amount in mdCATH is approximately 100 residues,  alanine dipeptide results in near-zero sequence similarity \cite{mirarchi2024mdcath}. Moreover, for more robust calculations, we umbrella-sampled the peptide to account for different conformations by applying a harmonic restraint force. To generate the free energy landscape, we utilized the Multi-Bennett Acceptance Ratio (MBAR) \cite{pymbar, shirts2008mbar}. After calculating the statistical MBAR weights, we applied a Gaussian smoothing function to approximate the free energy surface  as demonstrated in Figure 6.

The PHNN Ramachandran free energy landscape reproduces the major basins of alanine dipeptide with major caveats. The $\alpha_L$ basin ($\varphi \approx +60^\circ$, $\psi \approx +45^\circ$) is poorly represented in GBn2, which fails to stabilize this positive $\varphi$ region, while the $\alpha_R$ basin ($\varphi \approx -60^\circ$, $\psi \approx -45^\circ$) is poorly represented in PHNN with a significantly higher free energy than expected. The $\beta$/C$_{7eq}$ region ($\varphi \approx -120^\circ$, $\psi \approx +120^\circ$) is well reproduced by both models. While PHNN heavily distorts basin shapes, GBn2 tend to elongate regions of the plot. Nevertheless, PHNN's failure mode results from strained configurations and future iterations will explicitly account for conformational sampling. 
\subsection{Targeted analysis}

\begin{figure}[!h]
    \centering
    \includegraphics[width=1\linewidth]{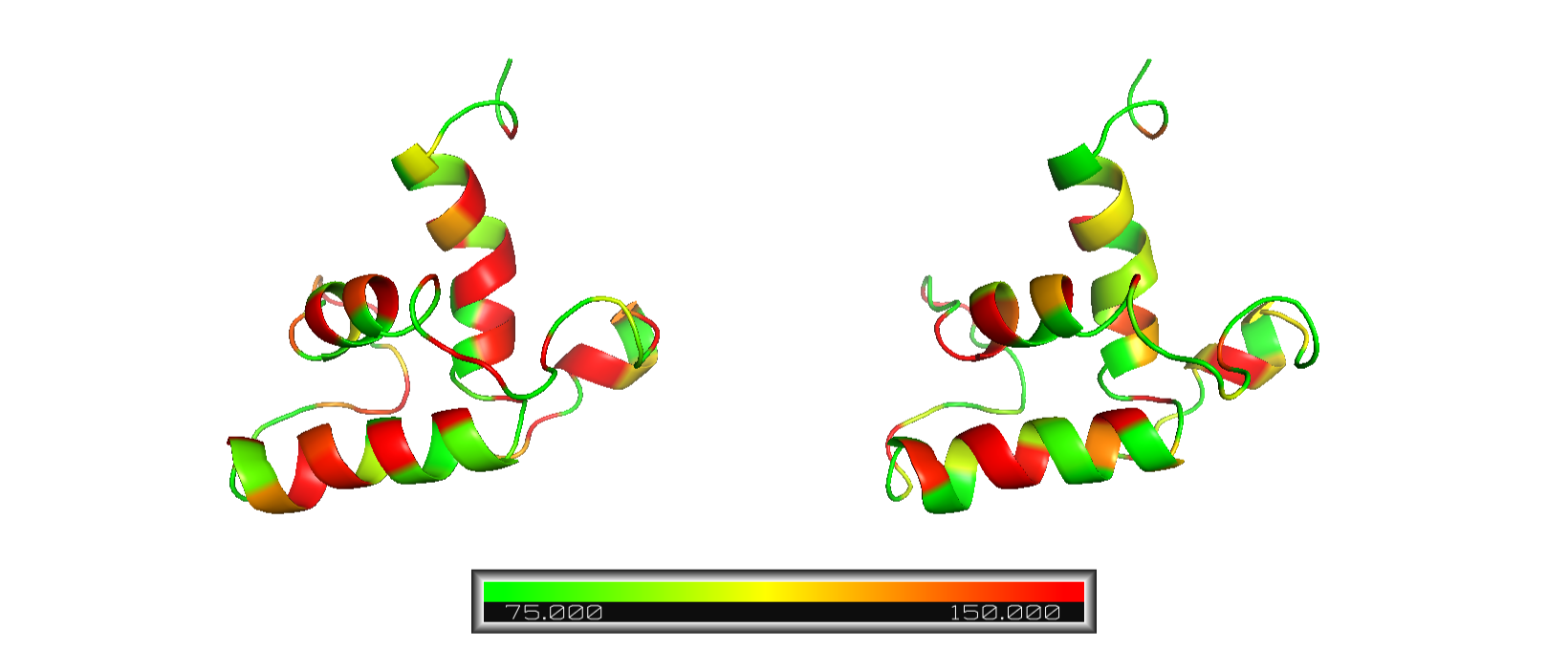}
    \caption{Mapping atomistic force error calculated on domain 3hb3B02 (1357 atoms) with a. GBn2 and b. PHNN}
    \label{fig:placeholder}
\end{figure}

While the information above provided a necessary overview of the model, a targeted analysis of specific targets will further assess the quality of PHNN. In Figure 7, a large improvement is visible on looping regions and alpha helices. The errors of PHNN are similarly consistent with existing errors in GBn2. 
\begin{figure}[!h]
    \centering
    \includegraphics[width=0.75\linewidth]{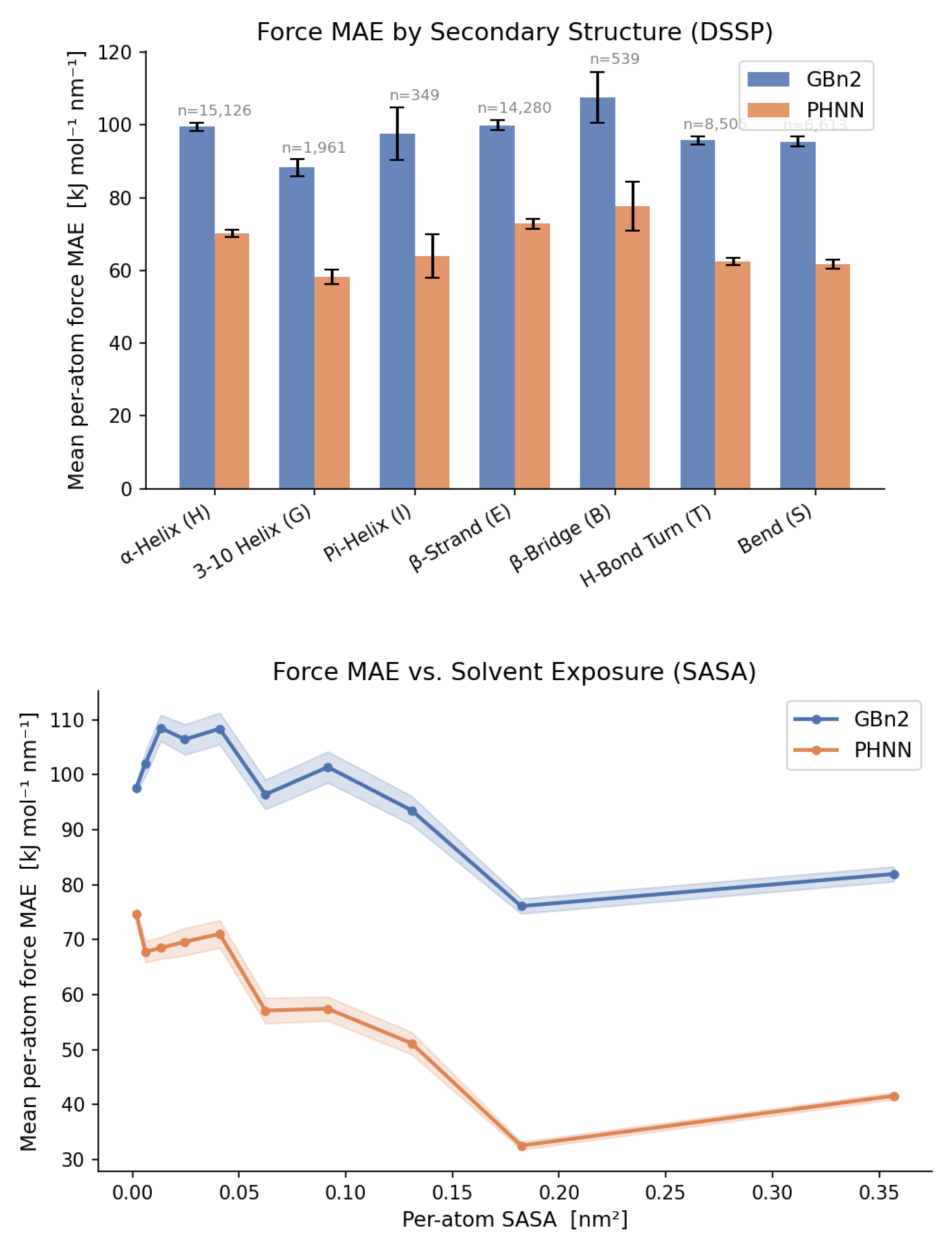}
    \caption{a. Force MAE by secondary structure calculated with DSSP; b. Force MAE vs. SASA}
    \label{fig:placeholder}
\end{figure}

A closer analysis is shown in Figure 8.  Across all secondary structures, PHNN consistently outperforms GBn2, with substantially lower MAE values.  PHNN’s largest improvements were found in 3-10 Helix (G) and Bend (S) regions, as expected. The improvement is especially pronounced in $\beta$-structures, such as $\beta$-Bridge and $\beta$-Strand, where GBn2 errors exceed 100 (kJ/(mol·nm))² while PHNN remains below 80 (kJ/(mol·nm))². Error bars are significantly larger on Pi-Helix (I) and $\beta$-Bridge (B); most likely because these regions are underrepresented in the dataset. Despite sample size differences, the overall trend clearly demonstrates that PHNN provides more accurate and consistent force predictions across diverse protein motifs.

PHNN achieves lower force error than GBn2 across protein atoms regardless of solvent exposure. However, the magnitude of this improvement depends strongly on solvent exposure. For buried atoms with low SASA, the performance gap between PHNN and GBn2 becomes significantly smaller. This suggests that both models perform similarly in the protein interior, where electrostatic screening is dominated by the protein dielectric environment and solvent contributions are minimal. This problem can most likely be solved with a deeper network or wider radius graph to better account for long-range electrostatics.

\begin{figure}[!h]
    \centering
    \includegraphics[width=1\linewidth]{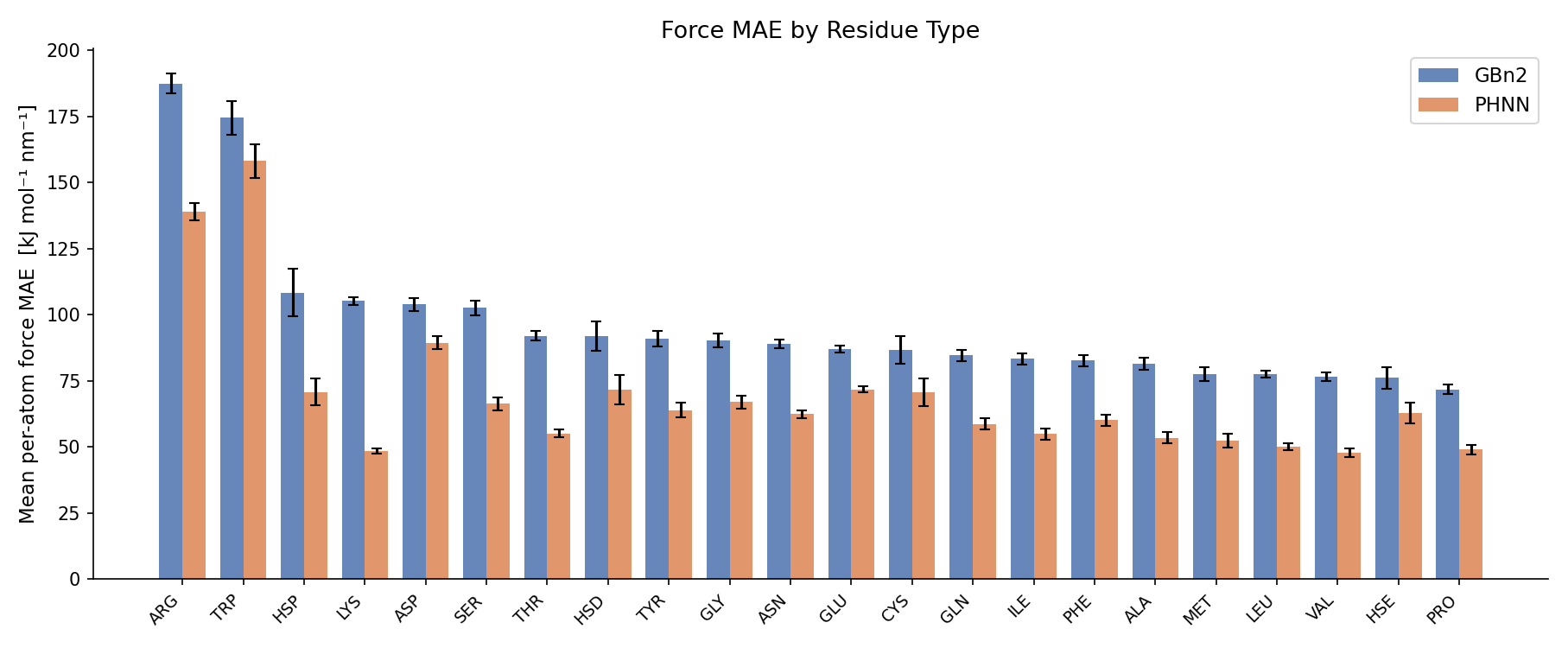}
    \caption{Force MAE by residue type}
    \label{fig:placeholder}
\end{figure}

Similar to the others, PHNN improves across every amino acid in Figure 9. PHNN has a marginal improvement for tryptophan; we suspect that TRP's indole ring has a large polarizable $\pi$ system that neither GBn2 nor PHNN captures well. PHNN’s largest improvement can be found in lysine, with a 54.02\% improvement, dropping to 48.43 ± 11.3 (kJ/(mol·nm))². As LYS and ASP/GLU are the canonical salt bridge partners in proteins, these results confirm that the learned screening function correction is working as intended. Despite PHNN's consistent improvements, ARG remains the highest error residue for both models, likely because its delocalized guanidinium charge, spread across multiple atoms, is difficult to screen correctly with a single per-atom correction. A unifying pattern is that both error magnitude and the degree of improvement correlate with charge state across residues. ASP, GLU, HSE, CYS, HSD, and TRP all improved by less than 22\%, suggesting PHNN handles the positive side much better than the negative side. In contrast, smaller amino acids such as THR, VAL, and SER all improve by more than 35\%. Despite proline having the lowest error with standard GBn2 with an MAE of 71.89 ± 3.2 (kJ/(mol·nm))², PHNN improves by 31.73\% with an MAE of 49.07 ± 2.7 (kJ/(mol·nm))², ranking third-lowest across all residues behind only lysine and valine.

\section{Discussion}

While PHNN remains competent as a valid implicit solvent model,it further demonstrates the necessity of a physical backbone  to guide neural potentials. While GBn2 is comparatively inaccurate, its failures correlate with specific conformations and folding patterns, rather than the sequence-identity-dependent underfitting seen in purely data-driven models. Throughout all experiments, a core design constraint of PHNN was data efficiency. While mdCATH is large in terms of conformations, the underlying protein diversity of approximately 5000 domains is modest relative to the diversity of known protein structures. The inclusion of E(3) equivariance allows the model to extract geometric information invariant to rotation and translation, reducing the amount of data required to learn physically meaningful representations. Similarly, the heteroscedastic loss discourages overfitting inherent in instantaneous forces. As such, PHNN proves to be better than GBn2 for proteins. 

\subsection{Future Direction}

The current iteration of PHNN displays preliminary results for the initial version of the model. As such, key improvements will be made. PHNN is currently trained for only 2 epochs, future iterations will train longer given regularization guarantees. Moreover, mdCATH was designed for simulations at 320K-450K, whereas most computational studies are conducted at 300K. The dataset also consists entirely of globular proteins. While transferability was largely successful, future work will explicitly target intrinsically disorder proteins (IDPs). Notably, mdCATH has announced plans to release 300K simulations, which future iterations of PHNN will incorporate to improve physiological relevance.

While the current model demonstrates transferability across folded protein topologies, training exclusively on near-native CATH configurations provides limited signal in the unfolded and extended regimes. As such, the current iteration of PHNN led to erroneous results at free energy boundaries. Future iterations will incorporate dipeptide umbrella sampling on dimer center-of-mass distances to enrich training data in these regions and produce accurate free energy barriers across the full conformational landscape \cite{charron2025cgschnet}.

PHNN's model architecture was limited by the necessity of speed and stability. For instance, the original coupling factor considered a coupling context provided by the equivariant layers. We decided not to move forward with this decision at this time, as it severely reduced the speed of force differentiation and destabilized the equivariant layers, preventing proper convergence. The coupling context also broadened the physical contributions of the neural network, and its removal prevents the model from re-learning physics already encoded in GBn2. Despite this, the model remains computationally intensive. Calculated from Figure 10, PHNN’s simulation speed is approximately 31 ms per frame on 1 L40S GPU, scaling with a time complexity of O(n). As explicit solvent models scale with a quadratic water box, PHNN may be faster with larger proteins within a single replica. For practical application, utilizing CUDA MPS, For practical application, PHNN was capable of running 21 simultaneous alanine dipeptide simulations with independent contexts using CUDA MPS. This number may increase with replicas in the same context. 

\begin{figure}[!h]
    \centering
    \includegraphics[width=1\linewidth]{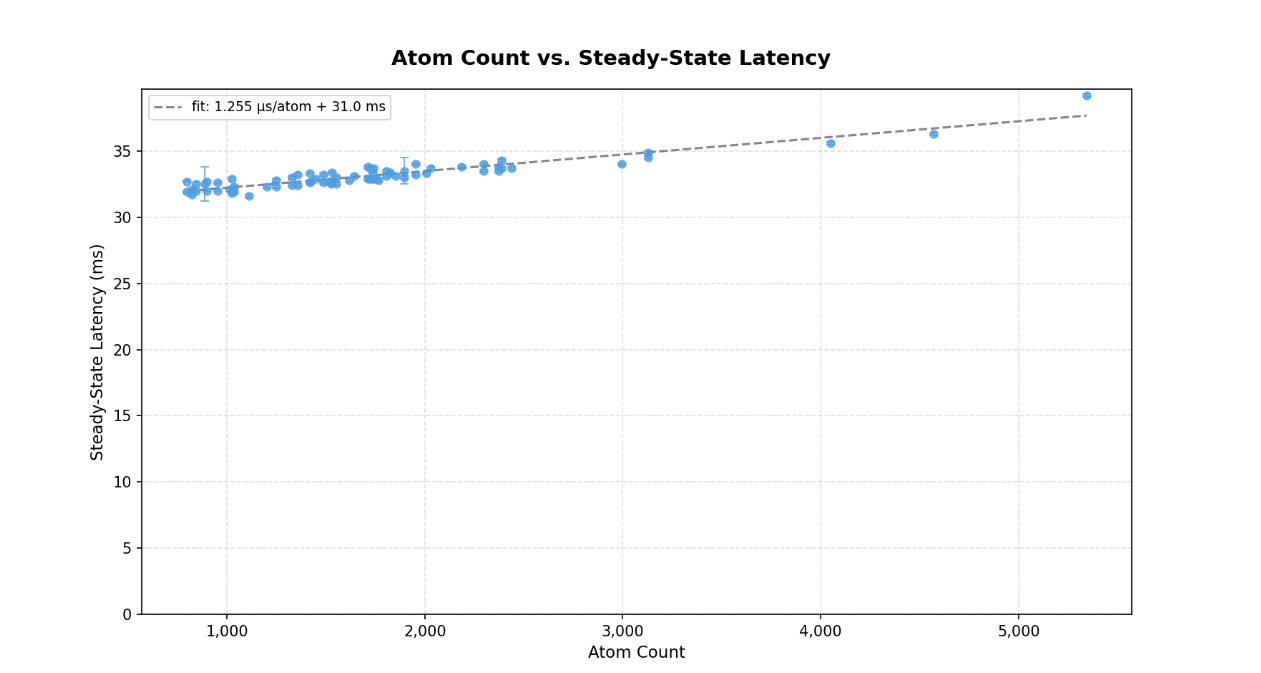}
    \caption{Averaged steady state latency over perturbed proteins coordinates (n = 39 proteins)}
    \label{fig:placeholder}
\end{figure}

Nonetheless, PHNN's increasingly accurate solvation model allows faster equilibration a period and conformational stability compared to GBn2. While our corrections are most pronounced where GB theory is known to be weakest, the accuracy gains are consistent across all folding patterns, confirming that PHNN's failure modes are architectural rather than protein-dependent. Buried region errors shared by both models, such as those seen in TRP, would require a deeper network or wider interaction radius to address, an improvement whose computational cost may not be justified given the minimal solvent contribution in the protein interior.

Lastly, we hypothesize that ARG’s high force errors are due to the inconsideration of the local charge environment. As discussed above, ARG's guanidinium charge delocalization suggests a per-atom correction is fundamentally insufficient for certain systems. Local charge density as explicit input to the screening function will be a cheap analytical solution and a natural architectural addition in a future modification.

\section{Conclusion}
PHNN successfully captures protein solvation while remaining transferable. While the current PHNN iteration was a proof of concept, future iterations will extend the training distribution to include dipeptide conformational sampling, enabling increased transferability to small molecule thermodynamics and folding free energy landscapes. Moreover, PHNN will contain a cleaner pipeline for direct application.

\bibliography{achemso-demo}

\end{document}